\documentclass[conference]{IEEEtran}
\IEEEoverridecommandlockouts
\usepackage{tikz}
\usepackage{pgfplots}
\usepackage{cite}
\usepackage{subcaption}
\usepackage{caption}
\usepackage{booktabs}
\usepackage{makecell}
\usepackage{setspace}
\usepackage{amsmath,amssymb,amsfonts}
\usepackage{graphicx}
\usepackage{hyperref}
\usepackage{textcomp}
\usepackage{multirow}
\usepackage[table]{xcolor}
\usepackage{xcolor}
\usepackage{epstopdf}
\usepackage{algorithm}
\usepackage{tabularx}
\usepackage{array} 
\usepackage{algpseudocode}
\usepackage{float}
\usepackage{placeins}
\usepackage{listings}
\usepackage{url}
\newcolumntype{Y}{>{\centering\arraybackslash}X} 

\usepackage{wrapfig}
\def\BibTeX{{\rm B\kern-.05em{\sc i\kern-.025em b}\kern-.08em
    T\kern-.1667em\lower.7ex\hbox{E}\kern-.125emX}}

\begin{document}

\title{Agentic Application in Power Grid Static Analysis: Automatic Code Generation and Error Correction\\
}

\author{\IEEEauthorblockN{Qinjuan Wang, Shan Yang, Yongli Zhu*}
\IEEEauthorblockA{Sun Yat-sen University, Guangzhou, China (wangqj35@mail2.sysu.edu.cn, yzhu16@alum.utk.edu)}
}

\maketitle

\begin{abstract}
This paper introduces an LLM agent that automates power grid static analysis by converting natural language into MATPOWER scripts. The framework utilizes DeepSeek-OCR to build an enhanced vector database from MATPOWER manuals. To ensure reliability, it devises a three-tier error-correction system: a static pre-check, a dynamic feedback loop, and a semantic validator. Operating via the Model Context Protocol, the tool enables asynchronous execution and automatically debugging in MATLAB. Experimental results demonstrate that the system achieves a 82.38\% accuracy regarding the code fidelity, effectively eliminating hallucinations even in complex analysis tasks.\\
\end{abstract}


\section{Introduction}
Static analysis, such as power flow and $N-1$ security assessment, is critical to the power grid's stable operation. To perform such routines, users (e.g., system operators, planners, university researchers) have to use professional software/tools, such as MATPOWER \cite{matpower_paper}, PSS/E, and so forth. Though most users are highly educated and experienced, coding every analysis task from scratch can be tedious and error-prone. Commercial-grade software, e.g, PSS/E and DIgSILENT, can support ``Scripting and Automation'' functionalities via a customized C++ or Python language subset. However, that still requires users to learn tautological programming manuals or grammar rules, which can be time-consuming. Moreover, debugging human-written code can be painful. As the modern power grid becomes increasingly complex, the above-mentioned limitations make it difficult for users to handle multiple \textit{ad hoc} system-analysis tasks in a timely manner.

A potential solution to overcome the above challenge is natural language processing (NLP)-based programming code generation, i.e., directly mapping a user's imperative instructions (in human language) to a script/code (in a target computer language). The emergence of Large Language Models (\textbf{LLMs}) and the related concept of \textbf{agents} or \textbf{agentic workflows} offer such possibilities.

For example, in \cite{projspecific}, the author proposes a method for inferring internal API information using RAG, which addresses the issue of inaccurate code completion by enabling LLMs to infer undocumented custom features. 
In \cite{codeact}, the author proposes CodeAct to integrate agent actions into a unified, executable Python space to solve complex problems through dynamic, multi-round interactions.
In \cite{anytool}, the author adopts AnyTool, a hierarchical agent that uses a self-reflection mechanism to reactivate the agent when the initial solution proves unfeasible. In \cite{knowledgefiltering}, the author proposes a knowledge filtering framework to eliminate noisy information and address the degradation in generation quality caused by irrelevant context.
In \cite{enable}, the author explores preliminary applications of LLMs in power system simulation, demonstrating the feasibility of LLM-based approaches for error feedback.

This paper presents an end-to-end LLM-based tool that automates the power grid static analysis from human users' natural language commands to executable scripting code for the corresponding tool, viz., MATPOWER, in this paper. The merits are: 1) RAG-based automatic script (code) generation, 2) mechanisms for error-correction, and 3) integration of advanced, industry-level toolchain(s) (e.g., DeepSeek-OCR \cite{deepseekocr}).

From the user's perspective, the proposed tool requires only a single natural language command to initiate the entire workflow; all subsequent code generation, execution, and iterative error correction are handled automatically by the agent. Internally, the agent employs a multi-turn feedback loop to resolve runtime errors and semantic inconsistencies.

Finally, the proposed tool is validated on a series of system analysis tasks of varying complexity, ranging from easy to hard. Accuracy indices are also defined to quantitatively evaluate our tool's performance.

\section{Problem Description}

\subsection{MATPOWER Introduction}

MATPOWER\cite{matpower_paper} is a toolbox designed for power grid static computation and optimization. The toolbox covers functionalities such as power flow (PF) (\textit{runpf},  optimal power flow (OPF) (\textit{runopf}), results visualization, and so on. It requires a structural input called ``case file" to specify bus-, line-, and generator-parameters, as well as other meta information.

\subsection{Scenarios when LLM-Agent is Needed}
For example, LLM-Agent-based automatic code generation is useful in the following scenarios:
\subsubsection{Online Contingency Analysis}
The power system contingency analysis is essentially a customized power flow computation that considers contingencies involving specific components. When the considered contingencies change (e.g., switching from branch to bus contingencies), the old program must be modified promptly.

\subsubsection{Operational Reliability Analysis}
The power system reliability analysis is essentially a Monte Carlo simulation combined with an (optimal) power flow calculation that considers combinations of component failures. When the considered failure types vary (e.g., from generator failure to load failure, the old program must be modified promptly. 

There are many other tasks that may benefit from LLM-agent-based automatic code generation (cf. Section \ref{sec4}).
). The next section describes the architecture of our agent tool.

\section{Automatic Code Generation and Error Correction}
\begin{figure}
    \centering
    \includegraphics[width=\linewidth]{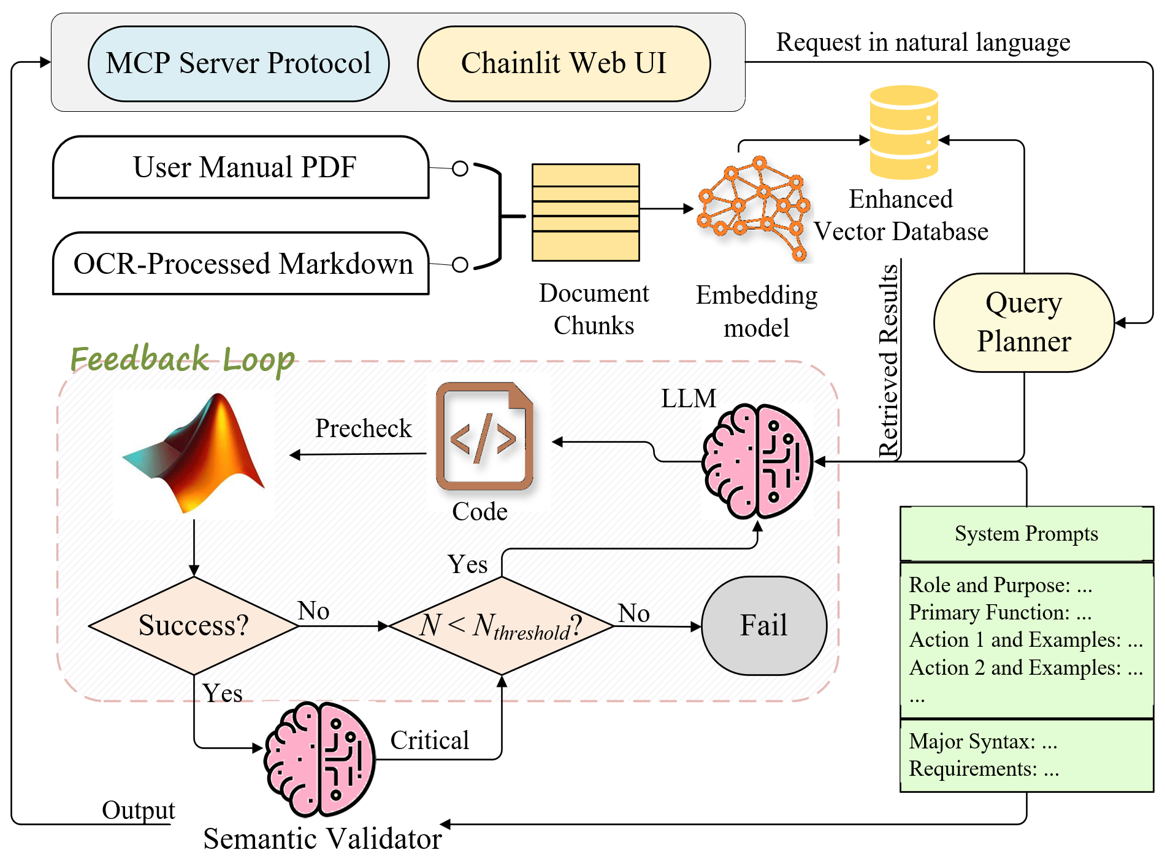}
    \caption{Overall architecture}
    \label{fig:architecture}
\end{figure}
\subsection{Software Architecture: MCP, RAG, and LangChain}
\subsubsection{LangChain Framework}
The system uses the LangChain framework\cite{langchain} as its brain and wraps the DeepSeek\cite{deepseek} interface. LangChain manages a long context window through system prompts, few-shot examples, and users' conversation history to equip the LLM with a role, ensuring its output follows the grammar and other constraints of MATLAB and the MATPOWER toolbox.
\subsubsection{Retrieval-Augmented Generation(RAG)}
To relieve the problems of the usual LLM, which lacks specific knowledge, such as power grid, and model hallucination, the system splits the user’s natural language instruction into multiple keywords through the query planner to analyze user intent before generating the code. It retrieves the relevant definition of API and its usage cases in the local knowledge base. This mechanism enhances the accuracy of code generation.
\subsubsection{Model Context Protocol (MCP)}
\begin{figure}[htbp]
    \centering
    \includegraphics[width=0.75\linewidth]{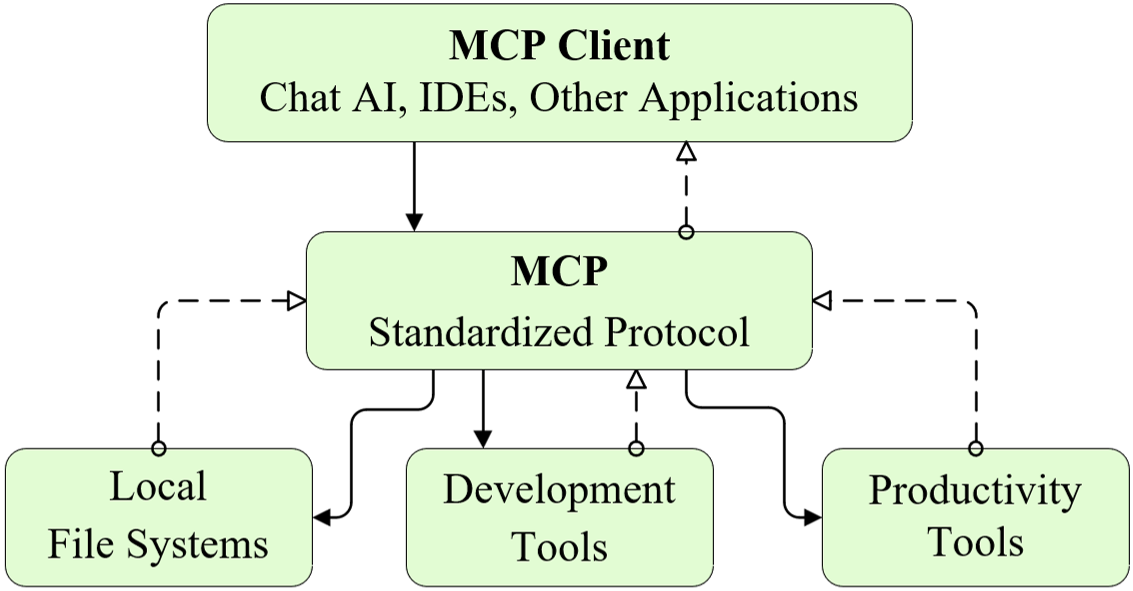}
    \caption{Model Context Protocol (MCP) Introduction}
    \label{fig:mcp_intro}
\end{figure}

The Model Context Protocol is a standardized communication protocol enabling AI agents to interact with external tools. Functioning similarly to a docking station, it extends their capabilities by providing access to diverse external resources(as shown in Fig. \ref{fig:mcp_intro}). The system implements the MCP standard \cite{mcp} and packages the entire function into an MCP server, enabling the system to run independently and to be integrated with AI agents that support MCP as a tool.

\subsection{Construction of the Vector Database and RAG Mechanism}
\subsubsection{Data Preprocessing Based on DeepSeek-OCR}
\begin{figure*}
  \centering
  \begin{subfigure}{0.32\linewidth}
    \centering
    \includegraphics[width=\linewidth,keepaspectratio]{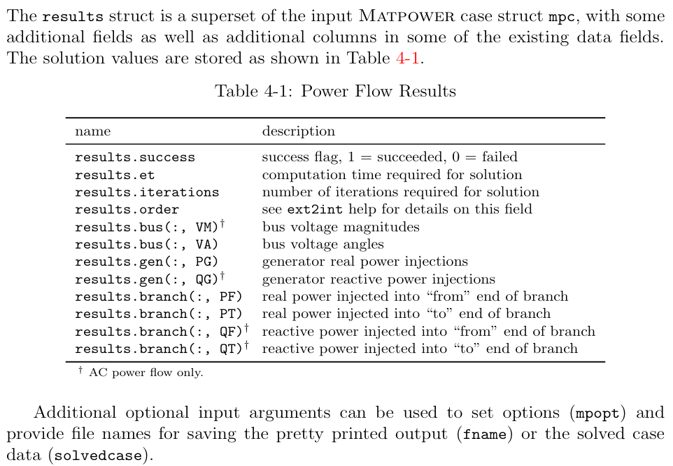}
    \caption{Raw PDF tables} 
    \label{fig:left}
  \end{subfigure}
  \begin{subfigure}{0.3\linewidth}
    \centering
    \includegraphics[width=\linewidth,keepaspectratio]{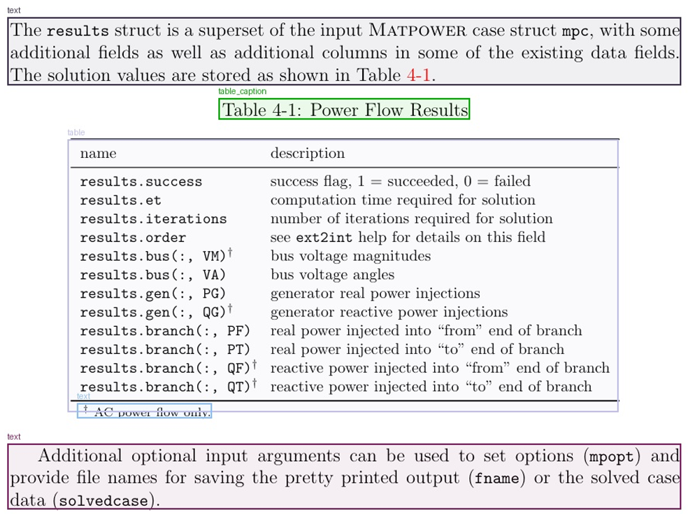}
    \caption{Identified authentic content} 
    \label{fig:left}
  \end{subfigure}
  \begin{subfigure}{0.31\linewidth}
    \centering
    \includegraphics[width=\linewidth,keepaspectratio]{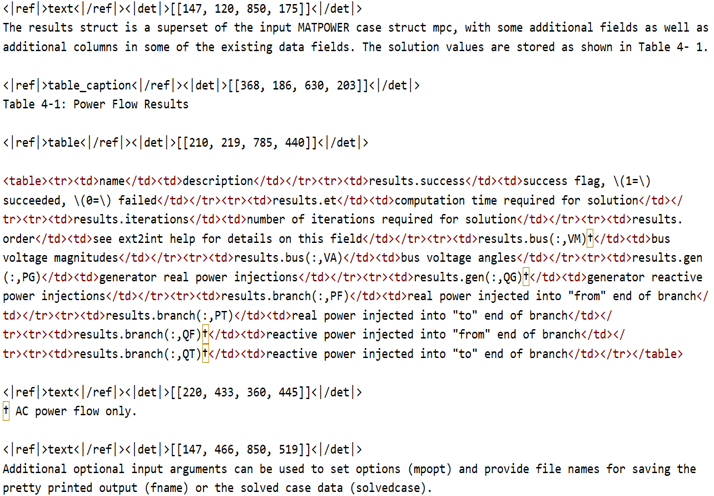}
    \caption{Structured Markdown output} 
    \label{fig:right}
  \end{subfigure}
  \caption{Example of DeepSeek-OCR processing}
  \label{fig:ocr}
\end{figure*}

The MATLAB toolbox user manuals often include complex tables of parameters, mathematical formulas, code examples, and hundreds of pages. However, conventional PDF text extraction tools only extract unformatted character streams and break the semantic logic of documents, which makes it difficult for LLMs to understand the relationships between functions and their parameters. To solve this problem, the system introduced DeepSeek-OCR \cite{deepseekocr} for document preprocessing. DeepSeek-OCR is an optical character recognition model designed for complex documents. It delivers outstanding performance in layout analysis and long-form content, and it can convert user manuals to Markdown with high quality.

Thus, the system processes the 265-page MATPOWER user manual \cite{matpower_manual} by running the DeepSeek-OCR model locally. The model precisely preserves the hierarchical headings and code blocks of the user manual, improving the accuracy of semantic segmentation during RAG retrieval. The system employs a fixed-size sliding-window algorithm to segment the markdown document, transforming it into fine-grained text chunks with independent semantics, laying the foundation for subsequent vectorization.

\subsubsection{Enhanced Vector Indexing Construction and Retrieval Strategy}\label{rag}
The system employs a pretrained transformer model to map the above text chunks to a high-dimensional semantic vector space, and constructs a vector indexing library using FAISS \cite{faiss}, a library for efficient similarity search and clustering of dense vectors.

When constructing enhanced vector indexing, the system utilizes multiple source data, concatenating OCR-processed results (as shown in Fig. \ref{fig:ocr}) with the original user manual text stream. This approach combines the strengths of both methods, ensuring the completeness of content and continuity of concepts while supplementing the row-column logic of complex tables and the syntactic structure of code blocks.

Within the overall framework, the RAG mechanism does not match user requests directly; instead, it decomposes them into sub-requests and maps each sub-request to its keywords precisely. Using these keywords, extract the top k most relevant fragments from the vector database. Then reassemble them and inject them into the LLM prompt as knowledge context to improve the usability of the generated code.

\subsection{Construction of the MCP/LangChain}
The following details the specific architectural implementation of the LangChain-based agent workflow and MCP server. The module serves as the core controller of the system, responsible for coordinating the LLM's cognition with interactions in the underlying executor environment.
\subsubsection{MATPOWER Agent Workflow}
After receiving the retrieval results (as mentioned in Section \ref{rag}), the agent proceeds to the prompt construction. Using the LangChain message management mechanism, it generates a composite system prompt that employs a three-layer structure. 

The top layer defines role parameters, specifying the coding language and constraints. The middle layer injects knowledge, e.g., the top-\textit{k} retrieved manual fragments and standard MATPOWER API conventions, while the bottom layer provides task instructions with user requests and few-shot examples. This structure ensures that the LLM’s reasoning is confined within the valid semantic space of the MATPOWER toolbox.

Then, the agent launches a MATLAB engine session through the MATLAB Executor module. Executes generated code in MATLAB, monitors execution status, captures warning and error messages, and handles any exceptions that may occur.

Specifically, the MATPOWER agent contains logic to drive the “generate-execute-correct” loop. When code execution fails in MATLAB, the agent does not report an error and exit immediately. Instead, it automatically triggers an error correction mode. The agent iterates this loop continuously until either the maximum threshold is reached or semantic validation succeeds.

\subsubsection{MCP Architecture and Inter-Process Communication}
\begin{figure}[htbp]
    \centering
    \includegraphics[width=\linewidth]{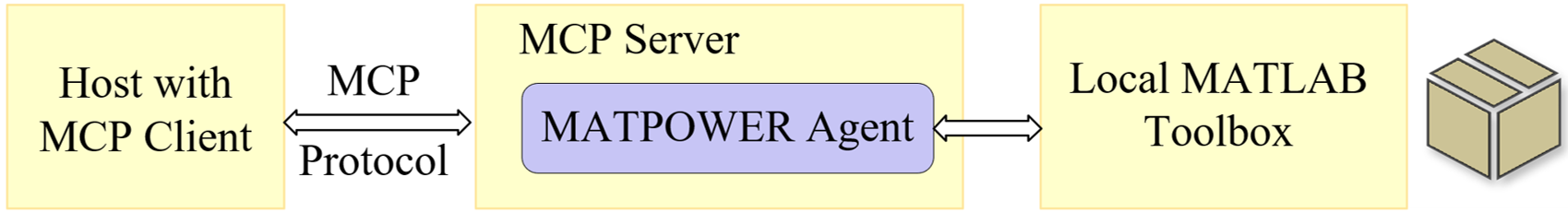}
    \caption{MCP architecture}
    \label{fig:mcp}
\end{figure}
To overcome the closed nature of conventional scripting tools, the system implements an MCP server that enables it to be invoked as a tool by the external ecosystem (Fig. \ref{fig:mcp}).

Since MATLAB simulations are computationally intensive, the MCP server does not use a simple synchronous approach. Instead, it uses an asynchronous architecture that combines asyncio and subprocess. This allows the MCP server to avoid blocking other simulation tasks while waiting for a simulation task to complete. When receiving a request, the server process forks an independent Python subprocess to run the MATPOWER agent, preventing the entire service from becoming unavailable if a single simulation hangs.

To address data exchange between the two processes, the system designs an application-layer protocol. After completing its task, the subprocess packages the final code, execution status, and debugging logs into a standard JSON data packet and sends it via the standard output stream. To precisely capture target data from a stream mixed with MATLAB logs, the subprocess appends an identifier to the JSON packet header. The MCP server monitors the subprocess’s output, captures and parses the subsequent JSON content immediately when recognizing the identifier, converts it into a format compliant with the MCP standard, and responds to the client.

\subsection{Mechanism of Error Feedback and Correction}
To address hallucinations and logical errors in code generated by LLMs, the system employs a three-tier architecture comprising a static pre-check, a dynamic feedback loop, and a semantic validator.
\subsubsection{Static Pre-check}
\begin{figure}
    \centering
    \includegraphics[width=0.7\linewidth]{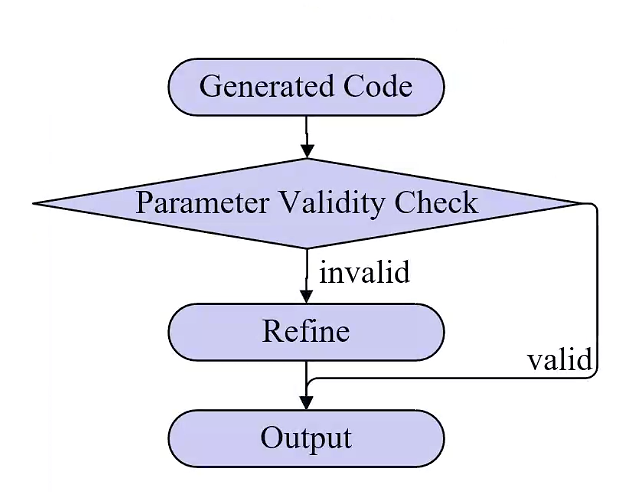}
    \caption{Static Pre-check}
    \label{fig:pre-check}
\end{figure}

Before MATLAB execution, the static pre-check module performs a scan based on MATPOWER conventions, as shown in Fig. \ref{fig:pre-check}. It utilizes fuzzy matching to rectify typos in option names and automatically injects the \textit{$define\_constants;$} statement if specific constants (e.g. \textit{$PD$} or \textit{$GEN\_BUS$}) are detected. This refinement filters out elementary errors, minimizing the overhead of the iterative feedback loop.
\subsubsection{Feedback Loop}
As shown in Fig. \ref{fig:architecture}, when MATLAB returns a runtime error, the agent starts the feedback loop. The system captures the error stack message and combines it with static pre-check hints to form an error report. This report also includes the failed code from the previous iteration, the user request, and related context, presented in RAG format, to construct new prompt feedback for the LLM. Then LLM analyzes and rewrites the code accordingly. This process sets a maximum iteration threshold to prevent falling into an endless loop.

\subsubsection{Semantic Validator}
To address cases where the generated code runs successfully but fails to meet the user's needs, the system adds a semantic sanity-check module after the MATLAB executor. An independent LLM-based semantic validator compares the user request against the final code. If logical inconsistencies are detected, they are considered semantic errors. Considering potential contradictions in the user request, the assessment results are quantified as “Critical” or “Minor” levels,
\begin{itemize}
    \item Critical-level semantic deviations are forcibly deemed failures, triggering a new round of the feedback loop.
    \item Minor-level semantic deviations are output as warning messages, ensuring generated scripts pass the check despite flaws in user requests.
\end{itemize}

\begin{figure}
    \centering
    \includegraphics[width=\linewidth]{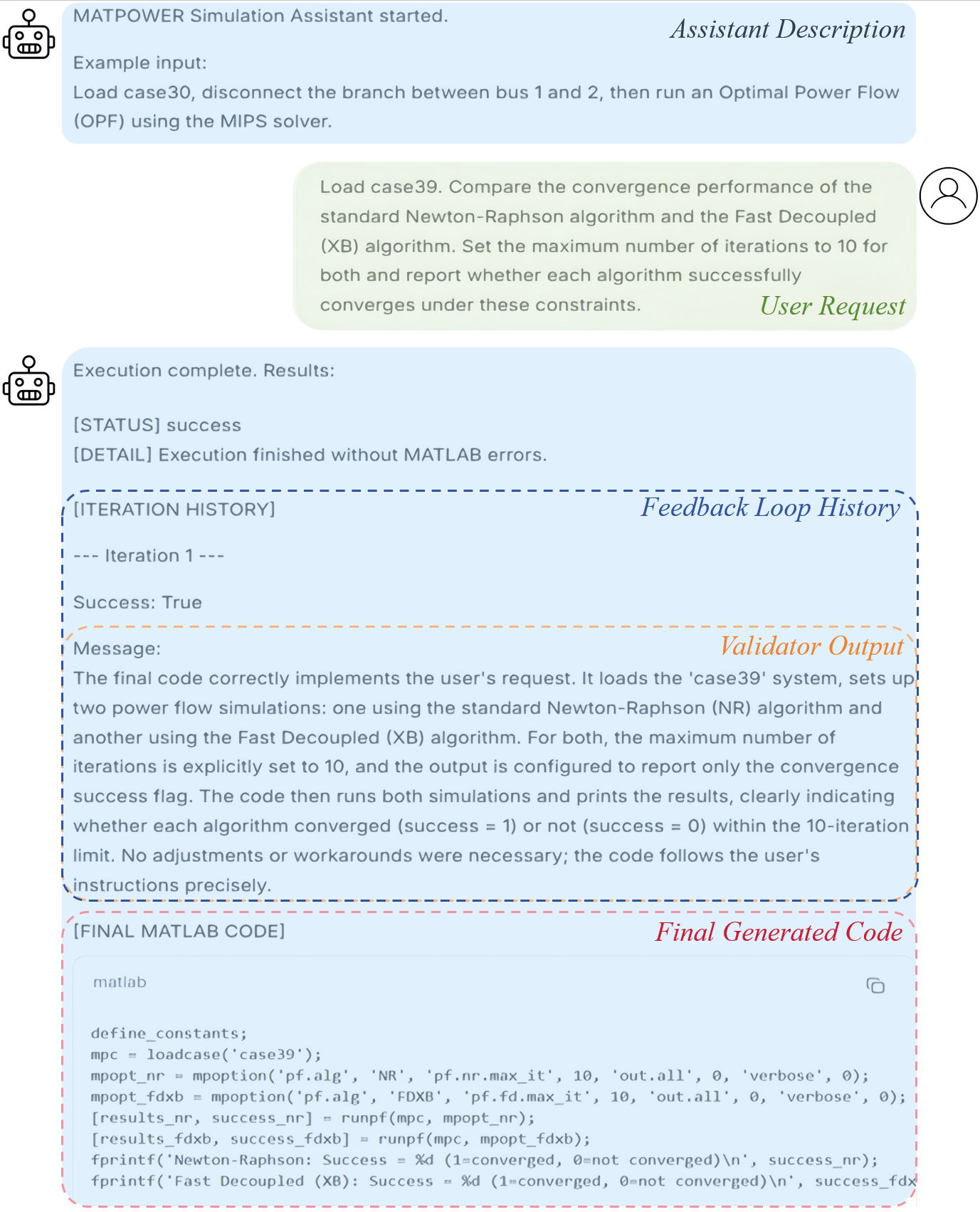}
    \caption{Output visualization}
    \label{fig:output}
\end{figure}
\subsection{GUI and Output Visualization}
To enhance the interactive experience, this paper also develops a web visualization using the \textit{Chainlit} framework \cite{chainlit}. Users can submit requests in natural language, and the interface provides real-time feedback on the system’s thinking process (as shown in Fig. \ref{fig:output}).

\section{Case Study}
\label{sec4}
To assess the performance of the proposed MATPOWER agent, basic and ablation experiments are conducted, focusing on retrieval strategies, the feedback loop, the query planner, and the semantic validator.

\subsection{Case Settings}
The basic experiment sets up the following four retrieval modes,
\begin{itemize}
    \item \textbf{no RAG}
    \item \textbf{Mode 1 (OCR-Markdown)} Perform keywords matching directly on the OCR-processed Markdown without constructing a vector database,
    \item \textbf{Mode 2 (PDF Vector DB)} Construct a vector database using the original PDF manual only, extracting original text fragments,
    \item \textbf{Mode 3 (Enhanced Vector DB, “RAG+”)} Merge PDF manual with OCR-processed Markdown to construct an enhanced vector database, combining semantic retrieval with structured information.
\end{itemize}

To investigate the effectiveness of each component in the proposed framework, the ablation experiment settings are shown in Table \ref{tab:ablation_config}.
\begin{table}[htbp]
  \centering
  \caption{Ablation Experiment: System Component Configurations}
  \label{tab:ablation_config}
  \setlength{\tabcolsep}{6pt}
  \renewcommand{\arraystretch}{1}
    {\fontsize{8pt}{9.6pt}\selectfont
  \begin{tabularx}{\columnwidth}{l*{4}{>{\centering\arraybackslash}X}}
    \toprule
    \multirow{2}{*}{\textbf{Ablation Configuration}} & \multicolumn{4}{c}{\textbf{RAG Mode 3}} \\
    \cmidrule(lr{0pt}){2-5}
    & \textbf{RAG+} & \textbf{Feedback} & \textbf{Planner} & \textbf{Validator} \\
    \midrule
    Full Model  & \checkmark & \checkmark & \checkmark & \checkmark \\
    Single Pass& \checkmark & & \checkmark & \checkmark \\
    Simple Search & \checkmark & \checkmark & & \checkmark \\
    Execution Only & \checkmark & \checkmark & \checkmark & \\
    \bottomrule
  \end{tabularx}
  }
\end{table}

These experiments test 10 simulation tasks, with a representative subset of six tasks detailed in Table \ref{tab:task}. These scenarios span a wide range of complexity, from standard power flow executions to sophisticated iterative algorithms and multi-matrix synchronization. Each task is executed in isolation to ensure results are free of interference from historical conversation context.

\begin{table}[htbp]
  \centering
  \caption{Representative Task Benchmarks}
  \label{tab:task}
  \setlength{\tabcolsep}{4pt}
  \renewcommand{\arraystretch}{1}
  {\fontsize{8pt}{9.6pt}\selectfont
  \begin{tabularx}{\columnwidth}{@{}llX@{}}
    \toprule
    \textbf{Task Type} & \textbf{Task ID} & \textbf{Natural Language Request} \\
    \midrule
    \multirow{3}{*}{Easy} 
    & Task 1 & Load case14. Increase the active load at bus 2 by 15\%. Run a DC power flow and display the results. \\
    & Task 2 & Load case57. Run a standard AC power flow and a DC power flow. Compare the resulting voltage magnitudes (Vm) and active power branch flows (Pf) between the two methods to evaluate the accuracy of the DC linearization. \\
    & Task 3 & Load case39. Compare the convergence performance of the standard Newton-Raphson algorithm and the Fast Decoupled (XB) algorithm. Set the maximum number of iterations to 10 for both and report whether each algorithm successfully converges. \\
    \midrule
    \multirow{7}{*}{Hard}
    & Task 1 & Load case39. Calculate the Total Transfer Capability (TTC) from Area 1 to Area 2. Write a MATLAB loop that iteratively increases Area 2 loads and Area 1 generation by 5\% increments. In each step, run a DC OPF and check if any branch flow exceeds its RATE\_A limit. Stop the iteration at the last feasible point before a violation and print the maximum successfully transferred power in MW. \\
    & Task 2 & Load case9. Add a new generator at Bus 4 (a load bus). You must change Bus 4 type to PV in mpc.bus, add the generator row to mpc.gen, and add a cost row to mpc.gencost. Ensure the BUS\_TYPE and generator location are perfectly synchronized before running OPF. \\
    & Task 3 & Run AC OPF on case30. Implement a user-defined linear constraint Pg1 + Pg2 $\le$ 30 MW using mpc.A, mpc.l, and mpc.u. Manually calculate the column indices for Pg1 and Pg2 based on the variable ordering without using helper functions. \\
    \bottomrule
  \end{tabularx}
  }
\end{table}

To quantitatively evaluate the agent's performance, we define the Code Generation Fidelity (CSGF) index. This metric considers both the semantic and the efficiency of the feedback loop.
\begin{equation}
    CSGF_i = S_{i} \times \left( \frac{N_{threshold} - (n_i - 1)}{N_{threshold}} \right)
\end{equation}
where $S_{i}$ is the semantic score for task $i$ (1.0 for perfect implementation, 0.8 for necessary technical workarounds, and 0 for final critical logic errors or final failure), $n_i$ is the number of iterations required to reach success, and $N_{threshold}=5$ is the maximum allowed iterations.

To assess the overall stability and reliability across the entire benchmark, the Global CSGF Accuracy (GCA) is calculated for each configuration as follows.
\begin{equation}
    GCA = \left( 1 - \sqrt{\frac{1}{K} \sum_{i=1}^{K} (1 - CSGF_i)^2} \right) \times 100\%
\end{equation}
where $K$ represents the total number of tasks.

\subsection{Experiment Results and Analysis}
The performance of the MATPOWER agent is comprehensively evaluated across diverse configurations using the CSGF and GCA metrics. 

\begin{figure}[htbp]
    \centering
    \includegraphics[width=\linewidth]{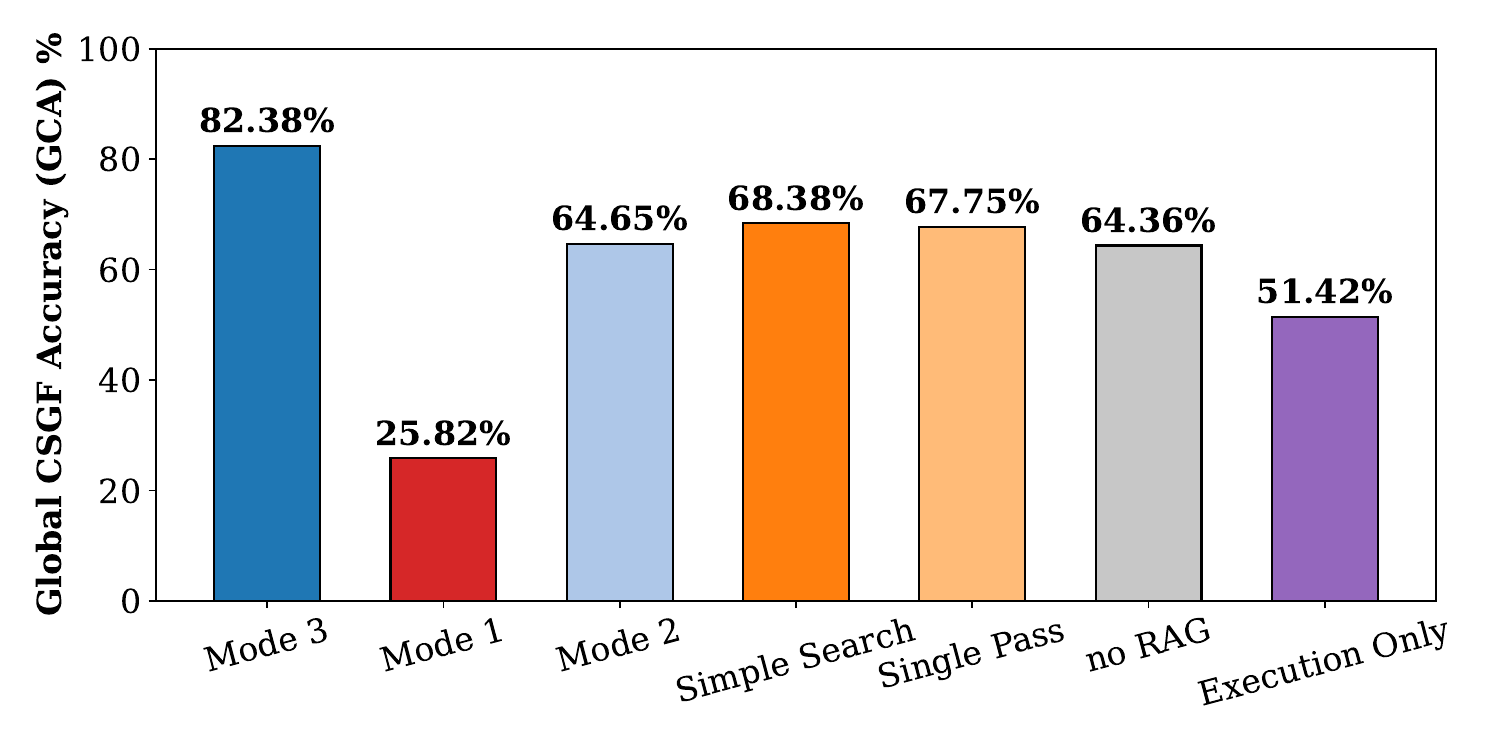}
    \caption{Overall system accuracy (GCA) across different configurations.}
    \label{fig:gca}
\end{figure}

The GCA for all configurations is illustrated in Fig. \ref{fig:gca}. The results show that the Full Model (Mode 3) is the most robust configuration, achieving a GCA of 82.38\%, which significantly outperforms all other retrieval modes and ablation variants.

Mode 2 utilizes traditional PDF chunking and achieves a GCA of 64.65\%. By integrating structured OCR-Markdown data, Mode 3 improved the generation fidelity by approximately 17.7\%. This validates that linearized, structured technical documentation provides a superior knowledge foundation for LLM-based agents compared to fragmented PDFs. 

\begin{figure}[htbp]
    \centering
    \includegraphics[width=\linewidth]{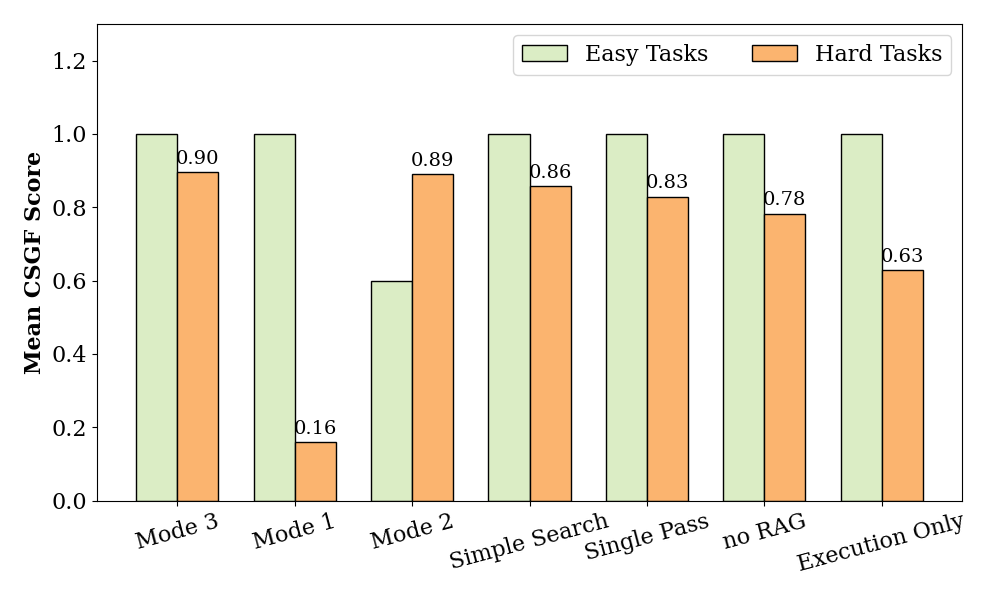}
    \caption{CSGF performance grouped by task complexity.}
    \label{fig:csgf}
\end{figure}

To investigate the agent's behavior under different logical loads, tasks are categorized into Easy and Hard groups. The mean CSGF scores per group are shown in Fig. \ref{fig:csgf}.

Mode 1 achieves a perfect score of 1.0 on easy tasks but plummets to 0.16 on hard tasks. This result indicates that while the OCR-processed Markdown document can meet simple requests, it fails to generate compliant code when handling complex logical operations, such as cross-matrix synchronous modifications (Hard Task 2) or manual variable index calculations (Hard Task 3), due to the absence of semantic associations.
Mode 2 underperforms on easy tasks compared to the Zero-shot, which is attributed to retrieval noise and context fragmentation inherent in raw PDF vectorization. Fragmented chunks may truncate simple API definitions or introduce irrelevant background information, distracting the LLM from straightforward instructions.

The ablation of the semantic validator (Execution Only) exhibits a deceptively high execution success rate, a manual audit of its logic and subsequent recalibration of the semantic score $S_i$ reveal a GCA collapse to 51.42\%. This confirms the agent's reliability. Without the validator, the agent produces about 31\% hallucinated code that is syntactically correct and executable in MATLAB but violates the user's intent. 

\section{Conclusion}
This paper addresses the challenges of laborious parameter configuration and debugging efforts in complex power system static analysis tasks by proposing an LLM-based automatic script generation framework that leverages enhanced retrieval and error correction. The experiment results demonstrate that the proposed framework significantly reduces hallucinations on hard tasks and the average number of iterations required for successful code execution.

The error-correction mechanism, combining static pre-check and dynamic feedback, enables the system to automatically repair errors in a finite number of iterations, even when the initially generated code contains errors. Although enhanced retrieval incurs slight delays in normal tasks, its advantage of avoiding multi-turn human-AI dialogues in complex tasks makes it more efficient than coding from scratch by humans. Future work will explore integrating our approach with other power system-related toolboxes.

\end{document}